\documentclass[final]{aipproc}

\layoutstyle{8x11double}

\usepackage{graphicx}
\usepackage{mdframed}
\usepackage{framed}
\usepackage{indentfirst}
\usepackage{natbib}
\usepackage{amsmath,amssymb}
\usepackage{subfigure}
\usepackage{enumitem}
\usepackage[margin=1in]{geometry}

\begin{document}

\title{\fontsize{18}{16.8} \selectfont Student Difficulties with the Dirac Delta Function}


\author{\fontsize{14}{15.6} \selectfont Bethany R. Wilcox}{
  address={Department of Physics, University of Colorado, 390 UCB, Boulder, CO 80309}
  }

\author{\fontsize{14}{15.6} \selectfont Steven J. Pollock}{
  address={Department of Physics, University of Colorado, 390 UCB, Boulder, CO 80309}
  }

\begin{abstract}
The Dirac delta function is a standard mathematical tool used in multiple topical areas in the undergraduate physics curriculum.  While Dirac delta functions are usually introduced in order to simplify a problem mathematically, students often struggle to manipulate and interpret them.  To better understand student difficulties with the delta function at the upper-division level, we examined responses to traditional exam questions and conducted think-aloud interviews.  Our analysis was guided by an analytical framework that focuses on how students activate, construct, execute, and reflect on the Dirac delta function in physics.  Here, we focus on student difficulties using the delta function to express charge distributions in the context of junior-level electrostatics.  Challenges included: invoking the delta function spontaneously, constructing two- and three-dimensional delta functions, integrating novel delta function expressions, and recognizing that the delta function can have units.  
\end{abstract}

\maketitle

\section{\label{sec:intro}Introduction}
\vspace*{-5pt}

Investigations aimed at identifying and understanding specific student difficulties with topics in physics are common both at the introductory and upper-division levels (see Ref.\ \cite{meltzer2012resource} for a review).  A key difference at the upper-division level is the increased importance of mathematical tools, making it less desirable to focus on conceptual and mathematical difficulties separately.    

One mathematical tool that students often encounter in their upper-division physics courses is the Dirac delta function (hereafter referred to as simply the delta function).  Delta functions are used in a variety of contexts throughout the physics curriculum including Fourier analysis, Green's functions, and as tools to express volume densities or potentials.  At the University of Colorado Boulder (CU), physics majors are usually introduced to the delta function in their middle-division classical mechanics course and encounter it again in both upper-division electrostatics and quantum mechanics.  

In the undergraduate curriculum, delta functions are often seen by experts as trivial to manipulate and are typically introduced to simplify the mathematics of a problem.  However, we have observed consistent student difficulties using the delta function.  This paper focuses on identifying student difficulties with the delta function in the context of electrostatics.  At CU, junior-level electrostatics is the first place where the delta function is embedded in a physical situation (e.g., to describe point, line, and plane charge densities).  Given the many uses of the delta function in various physics contexts, we do not claim that the issues we identify here span the space of student difficulties with the delta function; however, they do give us an idea of the kinds of challenges that students face when dealing with the Dirac delta function.  

This work is part of broader research efforts to investigate upper-division students' use of mathematics.  Problem solving at this level is often complex, thus an organizational structure is helpful to make sense of student difficulties.  We leverage the ACER framework (Activation, Construction, Execution, Reflection) \cite{wilcox2013acer} to scaffold our analysis of student difficulties with the delta function.  

ACER is an analytical framework that characterizes student difficulties with mathematics in upper-division physics by organizing the problem-solving process into four general components: \emph{activation} of mathematical tools, \emph{construction} of mathematical models, \emph{execution} of the mathematics, and \emph{reflection} on the results.  These components appear consistently in expert problem solving \cite{wilcox2013acer} and are explicitly based on a resources view on the nature of learning \cite{hammer2000resources}.  Since the particulars of how a mathematical tool is used in upper-division physics are often highly context-dependent, ACER is designed to be operationalized for specific mathematical tools in specific physics contexts.  Operationalization involves a content expert working through problems that exploit the targeted tool while carefully documenting their steps.  This process results in a researcher-guided outline of the key elements of a well-articulated and complete solution to these problems.  This outline is then refined based on analysis of student work (see Ref.\ \cite{wilcox2013acer} for details).

\vspace*{-14pt}
\section{\label{sec:methods}Methods}
\vspace*{-4pt}

Data for this study were collected from the first half of a two semester Electricity and Magnetism sequence at CU.  This course, E\&M 1, typically covers electrostatics and magnetostatics (i.e., chapters 1-6 of Griffiths \cite{griffiths1999em}).  The student population is composed of physics, astrophysics, and engineering physics majors with a typical class size of 30-60 students.  At CU, E\&M 1 is often taught with varying degrees of interactivity through the use of research-based teaching practices including peer instruction using clickers and tutorials \cite{chasteen2012transforming}.  

To investigate student difficulties with delta functions, we collected student work from three sources: traditional midterm exam solutions (N=303), the Colorado Upper-division Electrostatics Diagnostic (CUE, N=84), and two sets of think-aloud interviews (N=11).  Exam data were collected from five different semesters of CU's junior E\&M 1 course taught by four different instructors.  The only instructor to teach the course twice was a physics education researcher and the rest were traditional research faculty.  Interviewees were paid volunteers who had successfully completed E\&M 1 one or two semesters prior with one of three of these instructors, and who responded to an email request for participants.  

Questions on the exams and CUE diagnostic provided the students with the mathematical expression for a charge (or mass) density and asked for a description and/or sketch of the distribution (e.g., Fig.\ \ref{fig:deltaQs}\emph{(a)}).  The interviews were designed to explore the nature of preliminary difficulties identified in the exam solutions and thus, both interview protocols included questions like that in Fig.\ \ref{fig:deltaQs}\emph{(a)}.  Another goal of the interviews was to target elements of the Activation and Execution components that were not accessed by the exam and CUE data.  To do this, all interviews began with a description of the charge distribution and asked for a mathematical expression for the charge density (Fig.\ \ref{fig:deltaQs}\emph{(b)}).  The second set also ended by asking students to perform several context-free integrations of various delta function expressions (Fig.\ \ref{fig:e1}).  

Exams were analyzed by coding each element of the operationalized framework that appeared in the student's solution.  These elements were then further coded to identify fine-grained, emergent aspects of students' work.  Interviews were also analyzed by classifying each of the students' major moves into one of the four components of the framework.  As the CUE question was in a multiple-choice format, it provided quantitative data on the prevalence of certain difficulties.  

\begin{figure}
    \begin{minipage}{1\linewidth}
      \begin{mdframed}
      \begin{minipage}{.05\linewidth} 
        \fontsize{9}{10.8}\selectfont
        \hspace{-2mm}\emph{(a)} \vspace{6.5mm}
      \end{minipage}
      \begin{minipage}{.94\linewidth}
        \fontsize{9}{10.8}\selectfont 
        Sketch the charge distribution: $\rho (x,y,z) = c \delta (x-1)$\\ 
      \hspace{7mm} Describe the distribution in words too.  What are the units of the constant, c? \vspace{-1mm}
      \end{minipage}
      \end{mdframed}
      \vspace{-5mm}
    \end{minipage} 
\end{figure}
\begin{figure}
    \begin{minipage}{1\linewidth}
      \begin{mdframed}
      \begin{minipage}{.05\linewidth} 
        \fontsize{9}{10.8}\selectfont
        \hspace{-2mm}\emph{(b)} \vspace{10.5mm}
      \end{minipage}
      \begin{minipage}{.94\linewidth}
        \fontsize{9}{10.8}\selectfont 
        Provide a mathematical expression for the volume charge density, $\rho(\vec{r})$, of an infinite line of charge running parallel to the z-axis and passing through the point $(1,2,0)$.  Define any new symbols you introduce.\\ \vspace{-4mm}
      \end{minipage}
      \end{mdframed}
    \end{minipage} 
  \label{fig:deltaQs}
  \caption{Example questions that align with (a) element A1 and (b) element A2 of the ACER framework.}  
\end{figure}

\vspace*{-14pt}
\section{\label{sec:acer}ACER \& Delta Functions}
\vspace*{-4pt}

We have operationalized ACER for the use of delta functions to express the volume charge densities of 1, 2, and 3D charge distributions.  For example, the volume charge density of a line charge passing through the point (1,2,0) can be expressed as $\rho (\vec{r}) = \lambda\delta(x-1)\delta(y-2)$, where $\lambda$ is a \emph{unitful constant} representing the charge per unit length.  Expressing volume charge densities in this way is often necessary when working with the differential forms of Maxwell's Equations and can facilitate working with the integral forms of both Coulomb's Law and the Biot-Savart law.  The operationalization of ACER for this type of delta functions problem is described below.  The element codes are for labeling purposes only and are not meant to suggest a particular order, nor are all elements always necessary for every problem.

\begin{figure}
\begin{minipage}{1\linewidth}
  \begin{mdframed}
   \begin{minipage}{.32\linewidth}
    \fontsize{9}{10.8}\selectfont
    \begin{enumerate}[label=\alph*),align=left]
      \item \vspace{-0.5mm}\hspace{-3mm}$\int\limits_\infty^{-\infty} \delta(x)dx$ \vspace{-1mm}
      \item \hspace{-3mm}$\int\limits_\infty^{-\infty} x\delta(x)dx$ \vspace{-2mm}
    \end{enumerate}
  \end{minipage}
  \begin{minipage}{.67\linewidth}
    \fontsize{9}{10.8}\selectfont
    \begin{enumerate}[label=\alph*)]
      \setcounter{enumi}{2}
      \vspace{-3.5mm}\item $\int\limits_0^{10} [a\delta(x-1)+b\delta(x+2)]dx$ \vspace{-1mm}
      \vspace{2mm} \item $\iiint a\delta(r-r')r^2 sin(\theta)dr d\phi d\theta$ \vspace{-2mm}
    \end{enumerate}
   \end{minipage}
  \end{mdframed}
\end{minipage}
\caption{Context-free integrations in the second set of interviews to target element E1 of the ACER framework.}\label{fig:e1}
\end{figure}

{\bf Activation of the tool:}  The first component of the framework involves identifying delta functions as the appropriate mathematical tool.  We identified two elements in the form of cues present in a prompt that are likely to activate resources associated with delta functions.  
\vspace{-2mm}
\begin{enumerate}[label=A\arabic*:, align=left]
  \item The question provides an expression for volume charge density in terms of delta functions
  \item The question asks for an expression of the \emph{volume} charge density of a charge distribution that includes point, line, or surface charges
\end{enumerate}
\vspace{-2mm}
We include element A1 because, in electrostatics, delta functions are often provided explicitly in the problem statement, effectively short-circuiting Activation.  

{\bf Construction of the model:}  Elements in this component are involved in mapping the mathematical expression for the charge density to a verbal or pictorial representation of the charge distribution or vice versa.
\vspace{-2mm}
\begin{enumerate}[label=C\arabic*:, align=left]
  \item Relate the shape of the charge distribution to the coordinate system and number of delta functions
  \item Relate the location of the charges with the argument(s) of the delta function(s)
  \item Establish the need for and/or physical meaning of the unitful constant in front of the delta function
\end{enumerate}
\vspace{-2mm}
For problems that also require integration of the delta function (e.g., to find total charge from $\rho(\vec{r})$) there are an additional two elements in construction related to setting up this integral.  However, no students struggled to set up the relatively simple Cartesian integrals in this study.  As such, these two elements have not been included here.  
%\vspace{-2mm}
%\begin{enumerate}[label=C\arabic*:, align=left]
%\setcounter{enumi}{3}
%  \item Express a differential volume element consistent with the coordinate system
%  \item Select limits consistent with the differential volume element and region of interest
%\end{enumerate}
%\vspace{-2mm}

{\bf Execution of the mathematics:}  This component of the framework deals with elements involved in executing the mathematical operations related to the delta function.  Since this component deals with actually performing mathematical operations, these elements are specific to problems requiring integration of the delta function.  
\vspace{-2mm}
\begin{enumerate}[label=E\arabic*:, align=left]
  \item Execute multivariable integrals which include one or more delta functions
  %\item If necessary, manipulate the resulting algebraic expression into a form that can be readily interpreted
\end{enumerate}
\vspace{-2mm}
When the results of the integrals in E1 must be simplified for interpretation, Execution would include a second element relating to algebraic manipulation; however, none of the integrals included in this study elicited or required significant algebraic manipulation.  
%For sufficiently simple charge distributions, one can bypass element E1 by considering the physical meaning of the unitful constant (e.g., Q(spherical shell) = $\sigma*4\pi r^2$).  

{\bf Reflection on the result:} This final component includes elements related to checking and interpreting aspects of the solution, including intermediate steps and the final result.  While many different techniques can be used to reflect on a physics problem, the following two are particularly common when dealing with delta functions.  
\vspace{-2mm}
\begin{enumerate}[label=R\arabic*:, align=left]
  \item Check/determine the units of all relevant quantities (e.g., Q, $\rho$, the unitful constant)
  \item Check that the physical meaning of the unitful constant is consistent with its units and the units of all other quantities
%  \item Verify that the value of the charge in a region is consistent with expectations
\end{enumerate}
\vspace{-2mm}
While these two elements are similar, we consider element R2 to be a higher-level reflection task in that it is seeking consistency between the student's physical interpretation of the unitful constant and other quantities.

\vspace*{-14pt}
\section{\label{sec:results}Results}
\vspace*{-4pt}

This section presents the analysis of common student difficulties with the Dirac delta function organized by component and element of the ACER framework.  

{\bf Activation of the tool:}  Elements A1 and A2 of the framework are cues embedded in the prompt that can lead students to identify delta functions as the correct mathematical tool.  Element A1 short-circuits this process by providing the delta functions as part of the prompt.  Thus A1 type problems (e.g., Fig.\ \ref{fig:deltaQs}\emph{(a)}) provide little information about student difficulties recognizing when the delta function is appropriate.  A2 type problems (e.g., Fig.\ \ref{fig:deltaQs}\emph{(b)}) offer more insight into Activation as they do not provide or prompt the use of the delta function.  

None of the exams included A2 type questions, but this element was specifically targeted in the first of the two interview sets.  When presented with the question shown in Fig.\ \ref{fig:deltaQs}\emph{(b)}, only 2 of 5 interview participants suggested the use of delta functions.  The remaining three participants all expressed confusion at being asked to provide a volume charge density of a 1-dimensional charge distribution.  Two of these students attempted to reconcile this by defining an arbitrary cylindrical volume, $V$, around the line charge and using $\rho = Q/V$.  Later in the interview, when presented with the expression for this charge density in terms of delta functions, all but one of the interviewees correctly interpreted the expression as describing a line charge.  This suggests that even after completing a junior electrostatics course, many students may have difficulty recognizing when the delta function is the appropriate mathematical tool even when they are able to provide a correct physical interpretation of it.  

{\bf Construction of the model:} On the exam and CUE questions, the students were provided with an expression for the charge density and asked for a description or sketch of the charge distribution.  Here, students needed to connect the provided coordinate system and number of delta functions to the shape of the charge distribution (element C1).  For example, the charge density in Fig.\ \ref{fig:deltaQs}\emph{(a)} represents an infinite plane of charge.  Roughly one quarter of students' solutions (25\%, N=77 of 303) had an incorrect shape on the exams.  On the CUE diagnostic administered at the end of the semester, the fraction of students who selected an incorrect shape increased to slightly less than half the students (42\%, N=35 of 84).  The most common difficulty was misidentifying volume charge densities with 1 or 2 delta functions as point charges (53\%, N=41 of 77).  The drop-off in student success on the CUE indicates that students are not forming and/or maintaining a robust understanding of how delta functions relate to the shape of a charge distribution.  

To explore element C1 in a different way, some of the interviews provided a description of the charge distribution rather than a mathematical expression (Fig.\ \ref{fig:deltaQs}\emph{(b)}).  Here, students needed to use this description to choose an appropriate coordinate system and to determine the number of delta functions.  Of the eight interview students given this type of question, three were able to correctly express the line charge density as the product of two 1D Cartesian delta functions.  Four of the remaining five students used a single delta function whose argument was the difference between two vectors, i.e., $\rho \propto \delta(\vec{r}-\vec{r}')$ with $\vec{r}'=(1,2,z)$.  Three of these students also integrated their expression over all $z$ while describing the line charge as a continuous sum of point charges.  This finding, along with the frequency at which the exam students misidentified charge densities as point charges, suggests that our students may have a strong association between delta functions and point charges.  

Determining the location of the charge distribution (element C2) was not a significant stumbling block for students.  None of the interview students and just over a tenth of the exam students (13\%, N=38 of 303) drew an incorrect position for the distribution.  The most common errors were switching the signs of the coordinates (37\%, N=14 of 38, e.g., locating the plane in Fig.\ \ref{fig:deltaQs}\emph{(a)} at x=-1) or having the wrong orientation of line or plane distributions (37\%, N=14 of 38).  All questions in this study have dealt with delta functions in Cartesian coordinates, and it is possible that student difficulties with element C2 would be more significant for non-Cartesian geometries.  
 
The third element in construction relates to the need for a unitful constant in the expression for $\rho(\vec{r})$.  For the exam data, this constant is provided, and we would like our students to consider its physical meaning.  For example, in Fig.\ \ref{fig:deltaQs}\emph{(a)}, the constant $c$ represents the charge per unit area on the surface of the plane.  Roughly a quarter (26\%, N=48 of 186) of the exam students presented with an arbitrary constant spontaneously commented on its physical meaning and most of these (90\%, N=43 of 48) had a correct interpretation.  More than just this quarter of students may have recognized the constant's physical significance but did not explicitly write it down.  The interviews suggest that a students' interpretation of the constant can be facilitated or impeded by their identification of its units.  This dynamic will be discussed in greater detail in relation to the Reflection component (below).  

{\bf Execution of the mathematics:}  One exam question provided an expression for the charge density of three point charges and asked for $\int \rho(\vec{r}) d\tau$.  Roughly a quarter of the students (27\%, N=15 of 56) made significant mathematical errors related to the delta function while executing this integral (element E1).  The most common error (73\%, N=11 of 15) amounted to a variation of equating the integral of the delta function with the integral of its vector argument.  This difficulty was also implicit in one third (32\%, N=27 of 84) of the responses to the CUE.  

The second interview set (N=6) targeted the first element in Execution differently by asking students to perform the context-free integrations shown in Fig.\ \ref{fig:e1}.  Two students stated that the integral in part b) would be equal to $x$ without evaluating this expression at $x=0$, but none of the six participants had difficulty with the integrals in parts a) or c).  This level of success is somewhat surprising given that a quarter of the exam students struggled to execute integrals that, to an expert, are very similar.  One explanation may be that the $\delta^3(\vec{r})$ notation used on the exam was harder for students to deal with than the mathematically equivalent $\delta(x)\delta(y)\delta(z)$. Three of six interviewees also evaluated the $r$ integral in part d) as if the delta function was not there (i.e., $\int \delta(r-r')r^2 dr=\frac{1}{3} r'^3$), despite correctly executing parts a)-c).  Their verbal explanations indicated that the issue was the delta function rather than the spherical integrals.  These results again suggest that students' success at common delta function integrals may not transfer to more complex integrals.  

{\bf Reflection on the result:}  For the questions used in this study, one of the most powerful tools available for checking and interpreting the various delta function expressions is looking at units (elements R1 and R2).  When asked for the units of the given constant (e.g., $c$ in Fig.\ \ref{fig:deltaQs}\emph{(a)}), two thirds of the exam students (69\%, N=128 of 186) gave correct units.  We would also like our students to consider the physical meaning of this unitful constant (element C3), but it was often difficult to assess if they had done so on our exam questions.  However, a third of students (32\%, N=60 of 186) gave units that were inconsistent with the geometry they identified.  This pattern indicates that they either did not have an appropriate physical interpretation of this constant (elements C3) or failed to connect that interpretation to the units (element R2).  

The interviews offer additional insight into the connection between the units and physical interpretation of the constant.  When prompted to comment on units, 9 of 11 participants explicitly argued (incorrectly) that delta functions were unitless and thus, regardless of the geometry of the charge distribution, the units of the constant must be $C/m^3$.  This argument was often justified by the statement that the delta function was `just a mathematical thing' and thus did not have units.  Four of these students had previously expressed a correct physical argument for the units of the constant.  In each case, the student either abandoned their physical interpretation or were unable to reconcile these conflicting ideas.  Ultimately, 7 of these 9 students required help from the interviewer to convince themselves of the units of the delta function.

\vspace*{-14pt}
\section{\label{sec:discussion}Concluding Remarks}
\vspace*{-4pt}

This paper presents an application of the ACER framework to guide analysis of student difficulties with the Dirac delta function in the context of mathematically expressing charge densities in junior-level electrostatics.  We find that our upper-division students have difficulty; (1) activating delta functions as the appropriate mathematical tool when not explicitly prompted, (2) translating a verbal description of a charge distribution into a mathematical formula for volume charge density, (3) transfering their knowledge of how to integrate delta functions to more complex and novel integrals, and (4) determining the units of the delta function in order to reflect on or check expressions for the charge density.  

These findings have several implications for teaching and assessing the use of delta functions in electrostatics.  Instructors should be aware that the canonical delta functions questions rarely require a student to consider when delta functions are appropriate.  Furthermore, constructing a mathematical expression for the charge density is a more challenging task than interpreting that same expression.  Additionally, the belief that the delta function is unitless was a surprising prevalent and persistent difficulty that may be exacerbated by presenting the delta function as a purely abstract mathematical construct.  

The ACER framework provided an organizing structure for our analysis that helped us identify nodes in students' work where key difficulties appear.  It also informed the development of interview protocols that targeted aspects of student problem solving not accessed by traditional exams.  The difficulties identified in this paper represent a subset of students' difficulites with the Dirac delta function and may not include issues that might arise from its uses in contexts outside of electrostatics.  

This work was funded by the NSF (CCLI Grant DUE-1023028 and GRF under Grant No. DGE 1144083).

%\vspace*{-14pt}
%\begin{theacknowledgments}
%\vspace*{-4pt}
%This work was funded by NSF-CCLI Grant DUE-1023028 and an NSF Graduate Research Fellowship under Grant No. DGE 1144083.
%\end{theacknowledgments}

\vspace*{-10pt}
\bibliographystyle{aipproc}   % if natbib is available
\bibliography{master-refs-5_14_14}

\end{document}